\documentclass[12pt,preprint]{aastex}
\usepackage{epsfig}
\usepackage{emulateapj5}
\usepackage{onecolfloat}

\shorttitle{THE FINE STRUCTURE OF SIMULATED GALAXY DISKS}
\shortauthors{ABADI ET AL.}

\newcommand{\gsim} { \lower .75ex \hbox{$\sim$} \llap{\raise .27ex \hbox{$>$}} }
\newcommand{\lsim} { \lower .75ex \hbox{$\sim$} \llap{\raise .27ex \hbox{$<$}} }

\begin{document}

\twocolumn[

\title{
Simulations of Galaxy Formation in a $\Lambda$CDM Universe II: \\
The Fine Structure of Simulated Galactic Disks}

\author{Mario G. Abadi\altaffilmark{1} and Julio F. Navarro\altaffilmark{2}}
\affil{Department of Physics and Astronomy, University of Victoria,
    Victoria, BC V8P 1A1, Canada}

\author{Matthias Steinmetz\altaffilmark{3}}
\affil{Steward Observatory, 933 North Cherry Avenue, Tucson, AZ 85721, USA, and 
Astrophysikalisches Institut Potsdam, An der Sternwarte 16, D-14482 Potsdam, Germany }

\and

\author{Vincent R. Eke \altaffilmark{4}}
\affil{Physics Department, Durham University, South Road, Durham DH1 3LE, England}

\begin{abstract}
We present a detailed analysis of the dynamical properties of a simulated disk
galaxy assembled hierarchically in the $\Lambda$CDM cosmogony.  At $z=0$, two
distinct dynamical components are easily identified solely on the basis of the
orbital parameters of stars in the galaxy: a slowly rotating, centrally
concentrated spheroid and a disk-like component largely supported by
rotation. These components are also clearly recognized in the surface brightness
profile of the galaxy, which can be very well approximated by the superposition
of an $R^{1/4}$ spheroid and an exponential disk. However, neither does the
dynamically-identified spheroid follow de Vaucouleurs' law nor is the disk
purely exponential, a result which calls for caution when estimating the
importance of the disk from traditional photometric decomposition
techniques. The disk may be further decomposed into a thin, dynamically cold
component with stars on nearly circular orbits and a hotter, thicker component
with orbital parameters transitional between the thin disk and the
spheroid. Supporting evidence for the presence of distinct thick and thin disk
components is found, as in the Milky Way, in the double-exponential vertical
structure of the disk and in abrupt changes in the vertical velocity
distribution as a function of age.  The dynamical origin of these components
offers intriguing clues to the assembly of spheroids and disks in the Milky Way
and other spirals. The spheroid is old, and has essentially no stars younger
than the time elapsed since the last major accretion event; $\sim 8$ Gyr ago for
the system we consider here. The majority of thin disk stars, on the other hand,
form after the merging activity is over, although a significant fraction ($\sim
15\%$) of thin-disk stars are old enough to predate the last major merger
event. This unexpected population of old disk stars consists mainly of the tidal
debris of satellites whose orbital plane was coincident with the disk and whose
orbits were circularized by dynamical friction prior to full disruption. More
than half of the stars in the thick disk share this origin, part of a trend that
becomes more pronounced with age: nine out of ten stars presently in the old
($\tau \gsim \, 10$ Gyr) disk component were actually brought into the disk by
satellites. By contrast, only one in two stars belonging to the old spheroid are
tidal debris; the rest may be traced to a major merger event that dispersed the
luminous progenitor at $z\sim 1.5$ and seeded the formation of the spheroid. Our
results highlight the role of satellite accretion events in shaping the
disk---as well as the spheroidal---component and reveal some of the clues to the
assembly process of a galaxy preserved in the detailed dynamics of old stellar
populations.
\end{abstract}

]
\altaffiltext{1}{Observatorio Astron\'omico, Universidad Nacional de C\'ordoba and Consejo Nacional de Investigaciones Cient\'{\i}ficas y T\'ecnicas, CONICET, Argentina; abadi@uvic.ca}
\altaffiltext{2}{Fellow of CIAR and of the Alfred P. Sloan Foundation; jfn@uvic.ca}
\altaffiltext{3}{Packard Fellow and Sloan Fellow; msteinmetz@aip.de}
\altaffiltext{4}{Royal Society University Research Fellow; V.R.Eke@durham.ac.uk}

\section{Introduction}
\label{sec:intro}

Some of the observed properties of galactic disks seem at odds with many of the
``natural'' trends expected in hierarchically clustering models and present a
significant challenge to the current paradigm of structure formation on small
scales (see, e.g., Sellwood \& Kosowsky 2001, Navarro \& Steinmetz 2000, Moore
2001).  Qualitatively, the main difficulty lies in reconciling the early
collapse and eventful merging history characteristic of the buildup of galactic
dark matter halos with the many dynamical clues which point to a smooth assembly
of the luminous component of galactic disks. This difficulty has made it
necessary to postulate a substantial role for complex astrophysical processes in
order to overcome some of these trends and to bring models of hierarchical
assembly into agreement with observation.

The prime concern regards the fragility of centrifugally supported
stellar disks to rapid fluctuations in the gravitational potential
such as those stirred by mergers and satellite accretion events
(Toth \& Ostriker 1992, Quinn, Hernquist \& Fullagar 1993,
Velazquez \& White 1999).  As first computed in detail by T\'oth \&
Ostriker (1992), the constraints on merger events undergone by a thin
disk such as that of the Milky Way are very strict indeed. These
authors find that less than $10\%$ of the disk mass within the solar
circle could have been accreted in the past $5$ Gyr in the form of
clumps, limiting severely the role of merging in the recent mass
accretion history of the Milky Way.

Although these numbers might be relaxed somewhat by taking into account the
coherent response of a self-gravitating disk to the accretion event (Huang \&
Carlberg 1997, Walker, Mihos \& Hernquist 1996, Vel\'azquez \& White 1999),
there is broad consensus that the presence of a dominant, thin, cold stellar
disk points convincingly to a history of mass accretion where major mergers have
played a minor role. Indeed, since the fragile circular orbits of disk stars are
easily perturbed during mergers, the age of the oldest disk stars is often used
to estimate the epoch of the last major merger. In the case of the Milky Way,
stars as old as $\sim 14$ Gyr are found in the solar neighborhood, suggesting a
protracted (and perhaps episodic) history of star formation in the disk (Wyse
2000, Rocha-Pinto et al 2000, Liu \& Chaboyer 2000) and, consequently, a paucity
of merger events at odds with typical merging histories of dark halos in
cosmogonies such as the $\Lambda$CDM model.


The need for ordered and smooth settling of gas into thin galactic disks leads
to a few basic predictions of hierarchical scenarios that can be contrasted
directly with observation. For example, because angular momentum results from
torques operating before collapse, it correlates directly with the radius and
time of turnaround (Navarro \& Steinmetz 1997). Thus, material that turns around
later is expected to have higher angular momentum, with two major implications:
(i) disks are expected to be physically smaller in the past, compared with
systems of similar rotation speed identified at present, and (ii) disks form
from the inside out, as higher angular momentum material, which collapses later,
should settle preferentially in the outer regions of the disk (Mo, Mao \& White
1998).

Although overall these expectations are not grossly inconsistent with
current data, a number of well established observations are worryingly
difficult to accommodate within this general scenario.  One of these
worries concerns the origin of the complex vertical structure of
galactic disks. Current datasets suggest that most (if not all)
galactic disks are built out of two major dynamical components,
referred to generally as the ``thick'' and ``thin'' disks (Dalcanton
\& Bernstein 2002 and references therein). Although varying in importance
from galaxy to galaxy, the thick disk appears to be old ($\gsim 10$-$12$ Gyr in
the case of the Milky Way, Gilmore, Wyse \& Jones 1995), poor in metals
(Prochaska et al 2000), and of similar radial extent as the thin disk component
(Wyse 2000).

These trends have led to the popular conjecture that the Milky Way's
thick disk has its origin in an early thin disk of velocity and size
comparable to today's but ``thickened'' by the accretion of a
satellite $\sim 10$-$12$ Gyr ago (Gilmore, Wyse \& Norris
2002). Appealing as this idea may be from a dynamical standpoint, it
is rather unattractive in a hierarchical scenario, where the mere
existence of such large, fast-rotating disks at $z\sim 2$
(corresponding to roughly $\sim 12$ Gyr ago in the $\Lambda$CDM
scenario) counters the ``natural'' prediction of the model.  A related
difficulty concerns the presence of old disk stars in the vicinity
of the Sun, which might be taken to imply that even at high redshift
the thin disk already extended as far as the solar circle and has
remained essentially undisturbed dynamically since. Such a long period
of quiet dynamical evolution is difficult to reconcile with the active
merging expected at early times in the $\Lambda$CDM cosmogony, and
perhaps even with the presence of the Galaxy's spheroidal component,
which is typically ascribed to a relatively recent major merger event.

The discussion above illustrates the still unsettled account of the
origin of the Galactic disk(s) in the hierarchical formation paradigm,
and draws notice to the fundamental role that accretion events, be
they mergers or satellite disruptions, have played in determining the
structure of galactic disks. In this paper---the second in a series
analyzing the dynamical properties of galaxies simulated in the
``concordance'' $\Lambda$CDM paradigm---we examine these issues using
a numerical simulation with high numerical resolution. The simulation
is the same as described by Abadi et al (2002) in Paper I of this
series; that paper confronts the global dynamical and photometric
properties of the simulated galaxy with normal spirals, while the
present one concentrates on the multi-component nature of the stellar
disk and on its dynamical origin.

The paper is organized as follows. In section~\ref{sec:numexp} we
briefly discuss, for completeness, numerical details of the
simulation; \S~\ref{sec:results} discusses the identification of
distinct dynamical components in the simulated galaxy, as well as the
implications of these results in the context of the formation of the
thick and thin disks of the Milky Way. Finally, \S~\ref{sec:conc}
summarizes our main conclusions.

\section{The Simulation}
\label{sec:numexp}

A complete description of the numerical simulation analyzed in this
paper is presented in Paper I (Abadi et al 2002), where the interested
reader may find full details. For completeness, we provide here a
brief description of the numerical method and of the simulation.

\subsection{The Code}
\label{ssec:code}

The simulation described here was performed with GRAPESPH, a particle-based,
fully three-dimensional Lagrangian hydrodynamical code that combines the
flexibility and adaptability of the Smoothed Particle Hydrodynamics technique
with the speed of the special-purpose hardware GRAPE for computing gravitational
interactions between particles (Steinmetz 1996). The version used here includes
the self-gravity of gas, stars, and dark matter, hydrodynamical pressure and
shocks, Compton and radiative cooling, as well as the heating effects of a
photoionizing UV background (see Navarro \& Steinmetz 1997, Steinmetz \& Navarro
1999, for more details).

Star formation is handled in GRAPESPH by means of a simple recipe for
transforming gas particles in collapsing, dense regions into stars at
a rate roughly proportional to the local dynamical timescale of the
gas. After formation, star particles are only affected by
gravitational forces, but they devolve energy and mass to their
surroundings in a crude attempt to mimic the energetic feedback from
evolving stars and supernovae. A fraction, $\epsilon_v=0.05$, of this
energy is invested in raising the kinetic energy of gas neighboring
active star forming regions. These motions are dissipated by shocks on
a timescale longer than the local cooling time, allowing for lower
star formation efficiencies and longer effective timescales for the
conversion of gas into stars. Still, our choice of parameters results
in just a minor fraction of cooled, dense gas being reheated and
returned to intergalactic space in diffuse form. The star formation
history in the simulated galaxy thus roughly traces the rate at which
gas cools and collapses to the center of dark matter halos.

The mass devolved by star particles during feedback events is added to
gas particles surrounding young stars, and tracked in the code as a
crude measure of the metal content of the gas.  This ``metallicity''
is in turn inherited by the stars each gas particle spawns. By all
accounts, this treatment of metal enrichment is quite rudimentary and,
therefore, we restrict the use of stellar metallicities to the
modeling of the spectrophotometric contribution of each star particle
to the luminous output of a galaxy, as described by Contardo,
Steinmetz \& Fritze-von Alvensleben (1998).

\subsection{The Simulation}
\label{ssec:sim}

The simulation evolves a region that forms, at $z=0$, a galaxy-sized
dark matter halo in the low-density, flat Cold Dark Matter
($\Lambda$CDM) scenario (Bahcall, Ostriker \& Steinhardt 1999). This
model is fully specified by the following choice of cosmological
parameters{\footnote{ We express the present value of Hubble's
constant as $H(z=0)=H_0=100\, h$ km s$^{-1}$ Mpc$^{-1}$}}:
$\Omega_0=0.3$, $h=0.65$, $\Omega_{\rm b}=0.019 \, h^{-2}$, and
$\Omega_{\Lambda}=0.7$, where $\Omega_0$, $\Omega_b$, and
$\Omega_{\Lambda}$ represent the present-day contribution to the
mass-energy density of matter, baryons, and a cosmological constant,
respectively, in units of the closure density. The power spectrum is
normalized so that at $z=0$ the linear rms amplitude of mass
fluctuations in $8 \, h^{-1}$\,Mpc spheres is $\sigma_8=0.9$, and we
assume that there is no ``tilt'' in the initial CDM power spectrum.

At $z=0$ the dark matter halo under consideration has a circular velocity of
$V_{200}=134$ km/s and a total mass of $M_{200}=5.6 \times 10^{11} \, h^{-1} \,
M_{\odot}$, measured at the virial radius, $r_{200}=134 \, h^{-1}$ kpc, where
the mean inner density contrast (relative to the critical density for closure)
is $200$. This region was identified in a cosmological simulation of a large
periodic box ($32.5 \, h^{-1}$ Mpc on a side) and resimulated at higher
resolution, including the tidal field of the original simulation.  The
high-resolution region (an ``amoeba''-shaped region contained within a cube of
$3.4 \, h^{-1}$ comoving Mpc on a side) is filled at the initial redshift,
$z_i=50$, with the same number of gas and dark matter particles. The gas and
dark matter particle mass is $m_{g}=2.1\times10^6\, h^{-1}\, M_{\odot}$ and
$m_{dm}=1.2\times 10^7\, h^{-1}\, M_{\odot}$, respectively. We adopt a Plummer
softening scalelength of $0.5$ (physical) kpc for all gravitational interactions
between pairs of particles.

The baryonic mass of the final galaxy is $\sim 10^{11} M_{\odot}$,
equivalent to roughly $\sim 36,000$ gas particles (or more than
$100,000$ star particles).  This resolution enables a detailed study
of the dynamical and photometric properties of the simulated galaxy,
including the identification of different populations of stars
according to age and/or kinematics.  Galaxy luminosities are computed
by simply adding up the luminosities of each star particle, taking
into account the time of creation of each particle (its ``age'') and
its metallicity, as described in detail by Contardo, Steinmetz \&
Fritze-von Alvensleben (1998). Corrections due to internal absorption
and inclination are neglected, except for a temporary
wavelength-dependent dimming introduced to take into account the
gradual dispersal of the dust clouds that obscure the formation sites
of young stars (Charlot \& Fall 2000). Further details may be
consulted in Paper I (Abadi et al 2002).

\begin{figure*}[t]
\epsscale{1.75}
\plotone{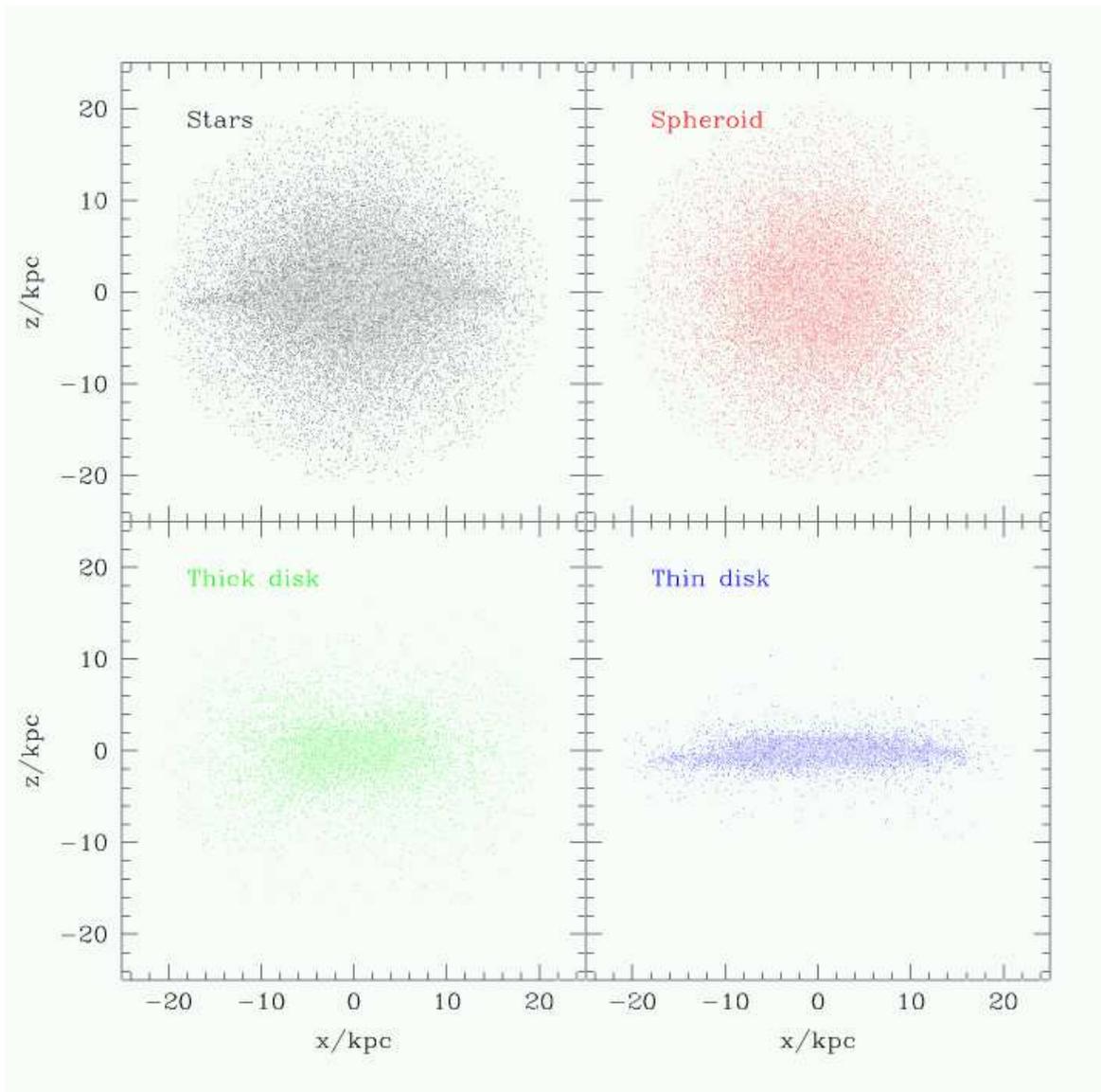}
\caption{Spatial distribution of star particles at $z=0$ projected so
that the disk component is seen edge on. Top-left: all
stars. Top-right: only those stars assigned to the spheroid by the
procedure outlined in \S~\ref{ssec:decomp} (see also
Figure~\ref{figs:je}). Bottom-left: same as top-right but for the
``thick disk'' component. Bottom right: same as top-right but for the
``thin disk'' component.}
\label{figs:stars}
\end{figure*}
\epsscale{1.0}

\section{Results} 
\label{sec:results}

The large-scale formation of the simulated galaxy is shown in Figures 1-3 of
Paper I, and follows the typical assembly process of dark matter halos in the
$\Lambda$CDM cosmogony. The region destined to form the halo collapses early
($z\lsim 5$) into a pancake-like structure criss-crossed by filaments along
which matter drains into collapsed dark matter halos. Gas cools and condenses at
the center of these halos, turning swiftly into stars as it settles at the
bottom of these potential wells. These progenitors gradually merge into a single
system at $z=0$ which is the subject of our analysis here.

The mass accretion history is punctuated by a succession of major
mergers at high redshift, although by $z\sim 1$ the merging activity
of the galaxy is over and the dark halo evolves little thereafter. The
cooled (and cooling) gas within the halo, on the other hand, keeps
flowing towards the center and settles gradually into a dynamically
cold, centrifugally supported disk. The stellar component of the
central galaxy increases by $\sim 40\%$ in mass since $z \sim 1$, and
is shown at $z=0$ in the top-left panel of Figure~\ref{figs:stars}.
Much of the late accreting material is clearly visible in the final
galaxy as a well-defined disk of young stars surrounding a more
centrally concentrated spheroid of older stars.  Photometrically, this
structure is reminiscent of that of early-type spirals and was
analyzed in detail in Paper I. We concentrate here on the detailed
dynamical analysis of the galaxy at $z=0$ and on the origin of its
major components.

\subsection{Dynamical Decomposition}
\label{ssec:decomp}

\begin{figure*}[t]
\epsscale{1.7}
\plotone{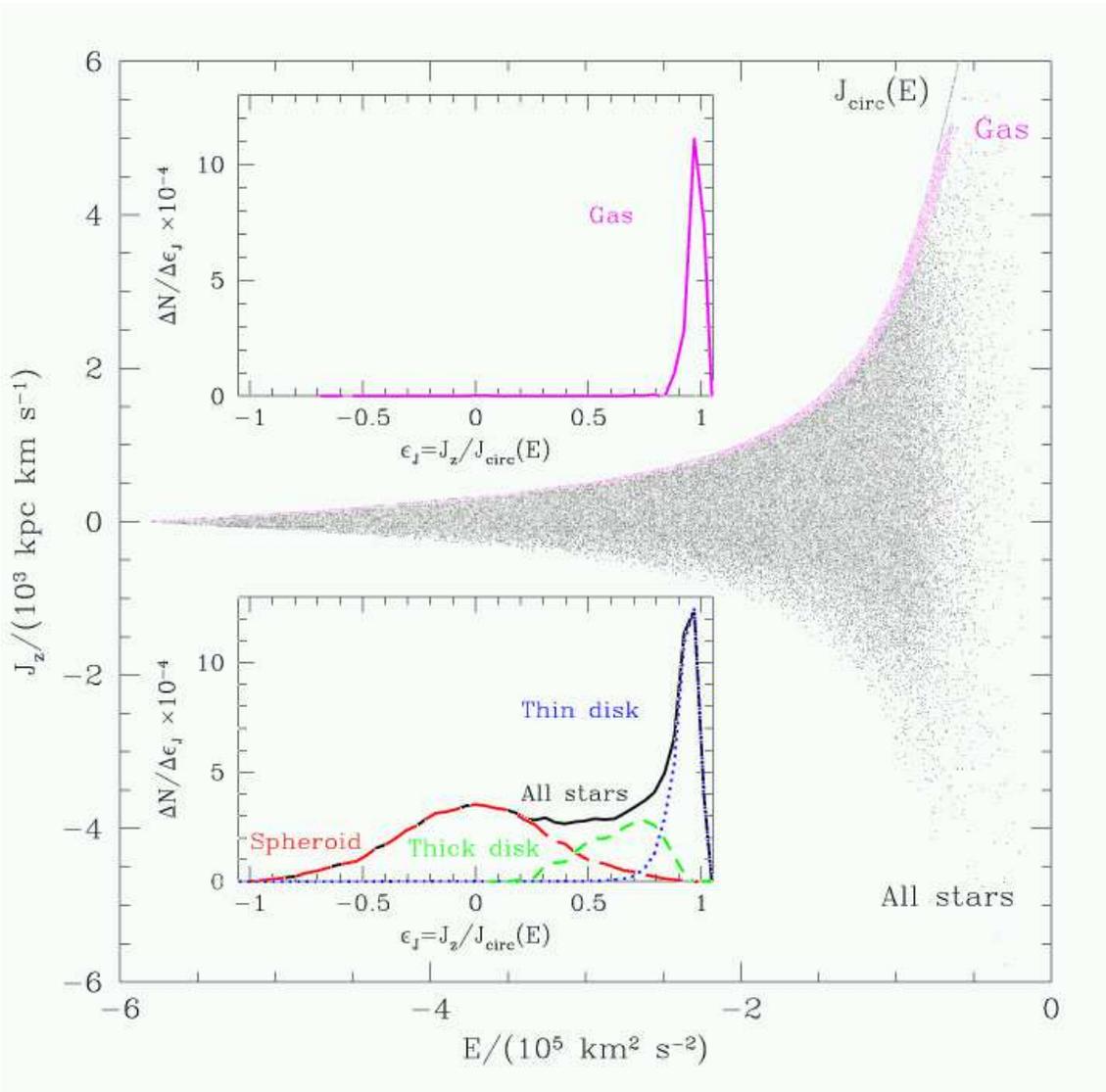}
\caption{The $z$-component of the specific angular momentum shown as a
function of specific binding energy for all star and gas particles within $21$
kpc from the center of the galaxy at $z=0$. The curve labelled $J_{\rm circ}(E)$
indicates the location of circular orbits co-rotating with the disk. The gas
particles are all on nearly circular orbits and follow very closely the $J_{\rm
circ}(E)$ curve, as shown by the magenta dots and by the distribution of the
circularity parameter $\epsilon_J=J_z/J_{\rm circ}$ shown in the top inset. The
bottom inset shows the circularity distribution of the stellar component, and a
possible decomposition into three different dynamical components: a spheroid
(long-dashed curve), a thin disk (dotted curve) and a thick disk (short-dashed
curve).  }
\label{figs:je}
\end{figure*}

The existence of a well defined, thin disk in the stellar and gaseous
components of the galaxy (see top-left panel of Figure~\ref{figs:stars})
defines unambiguously a symmetry axis which we shall hereafter call
the ``$z$-axis'', for short. Figure~\ref{figs:je} shows $J_z$, the
$z$-component of the specific angular momentum of all stars within the
``luminous'' radius of the central galaxy, $r_{\rm lum}=21$ kpc,
plotted versus the specific binding energy of each star, $E$. Binding
energies are computed relative to the total mass of the system within
the virial radius, $r_{200}$.  Particles with positive or negative
$J_z$ circulate around the center of the galaxy in different
directions; we choose $J_z$ to be positive so that most stars are
``co-rotating'' in Figure~\ref{figs:je}.

The available parameter space in the $J_z$-$E$ plane is delineated by
the maximum and minimum angular momentum consistent with a given value
of the specific energy: those of the co- and counter-rotating circular
orbits, respectively. The co-rotating circular orbits are shown by the
solid curve labelled $J_{\rm circ}(E)$ in Figure~\ref{figs:je}. The
gas particles within $r_{\rm lum}$ are on a thin, centrifugally
supported disk, and follow closely the $J_{\rm circ}(E)$ curve, as
shown by the magenta dots in Figure~\ref{figs:je}.  The prevalence of
circular orbits in the gaseous disk can also be seen in the
distribution of the ``orbital circularity'' parameter,
$\epsilon_J=J_z/J_{\rm circ}(E)$, which is sharply peaked about unity,
as shown in the top inset of Figure~\ref{figs:je}.

\epsscale{1.0}

The circularity distribution of stars also shows a sharp peak at
$\epsilon_J \sim 1$, but is quite broad and has approximately $25\%$
of stars on counter-rotating orbits (see solid line labelled ``all
stars'' in the bottom inset of Figure~\ref{figs:je}). The shape of the
distribution suggests the presence of at least two distinct dynamical
components; a hot spheroid with as many stars in co- as in
counter-rotating orbits, and a cold disk of stars with circularities
resembling those of the gas and sharply peaked about unity. The curves
labelled ``spheroid'' (long-dashed) and ``thin disk'' (dotted)
indicate a possible decomposition into such components. This
decomposition assumes that the spheroid has little net rotation, and
that the circularity distribution of the thin disk is similar to that
of a cold disk such as the thin disk of the Milky Way. Indeed, the
circularity distribution of the ``thin disk'' identified in the bottom
inset of Figure~\ref{figs:je} is similar to that of a $\sigma/V_{\rm
rot} \approx 0.2$ disk, where $\sigma$ and $V_{\rm rot}$ are the
velocity dispersion and the rotation speed of the disk, respectively.

The top and bottom right panels of Figure~\ref{figs:stars} show the
spatial distribution of stars selected according to the circularity
distributions labelled ``spheroid'' and ``thin disk'' in
Figure~\ref{figs:je}, respectively. These panels confirm the intuitive
expectation that the spheroid ought to be roughly spherical in shape
whereas the thin disk is highly flattened, with an aspect ratio
exceeding $10$:$1$. 

The decomposition into spheroid and thin disk proposed here leaves,
however, a number of stars on prograde orbits unassigned to either
component (see the short-dashed curve labelled ``thick disk'' in
Figure~\ref{figs:je}). These stars share the same sense of rotation as
the thin disk but rely less on rotation for their support; as a
result, they are distributed spatially on a ``thick disk'' shown in
the bottom left panel of Figure~\ref{figs:stars}.

It must be recognized that the identification of three distinct
dynamical components, although compelling in light of the shape of the
stellar circularity distribution, is nevertheless somewhat
arbitrary. One may question, for example, our assumption that the
spheroid does not rotate, or our assignment of essentially all stars
on circular orbits to the thin disk (realistic models for the thick
disk would typically have a sprinkling of such orbits as well). We
have experimented with several possible decomposition procedures and
report here results which, although they depend quantitatively on the
exact definition adopted for each component, are qualitatively
independent of such details and hold for every plausible decomposition
procedure. We list the main parameters of each component in
Table~\ref{tab:gxprop} and concentrate below on their origin as well
as on their imprint on the photometric properties of the galaxy.

\subsection{Photometric vs Dynamical Components} 
\label{ssec:photdyn}

\begin{figure}[t]
\plotone{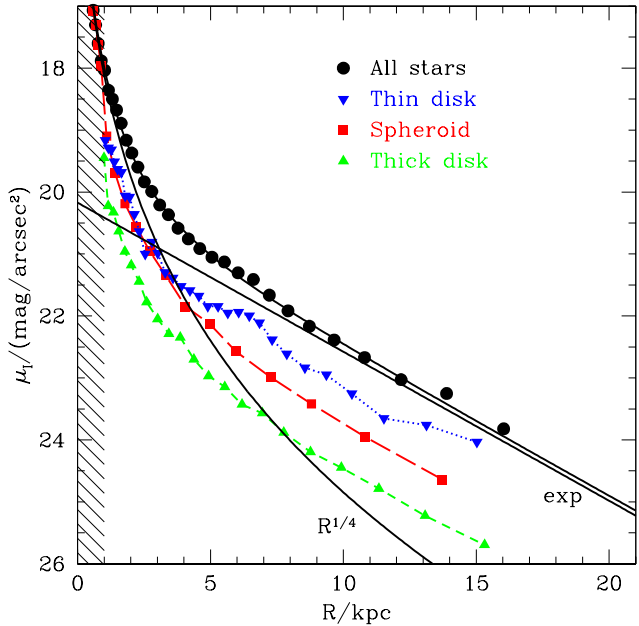}
\caption{$I$-band surface brightness profile of the simulated galaxy
seen face on. The solid curves show the best $R^{1/4}$+exponential
fit; triangles, inverted triangles and squares show the contribution
to the profile of the thick disk, the thin disk, and the spheroid,
respectively. Note that although the $R^{1/4}$+exponential fit
approximates very well the surface brightness profile, neither is the
disk exponential nor the spheroid $R^{1/4}$.  }
\label{figs:surfb}
\end{figure}

As discussed in Paper I, the surface brightness profile of the
simulated galaxy resembles that of early-type spirals, and can be
described very well by combining a central $R^{1/4}$ spheroid with a
surrounding exponential disk. This is shown by the solid circles in
Figure~\ref{figs:surfb}, where we show the $I$-band surface brightness
profile of the simulated galaxy seen face-on. The solid line through
the circles corresponds to the best $R^{1/4}$+exponential fit, whereas
the curves labelled ``$R^{1/4}$'' and ``exp'' indicate the
contribution of each of these components to the fit. The best
$R^{1/4}+$exponential fit splits the total light in the $I$ band
roughly equally between the disk and the spheroid, as in early type
spirals such as UGC615.  The virtues and shortcomings of this
comparison are discussed in detail in Paper I. Here we focus on the
dynamical interpretation of the fitting procedure.

Does the photometric decomposition describe accurately the true
dynamical importance of each component? Is the disk truly exponential?
Is the spheroid really well approximated by de Vaucouleurs' law? Our
simulation offers intriguing clues to these questions. The inverted
triangles and filled squares in Figure~\ref{figs:surfb} show the
contribution of the dynamically-identified thin disk and spheroid to
the face-on $I$-band surface brightness profile, respectively. 

The spheroid contributes most of the $I$-band light ($\sim 60\%$); the
thin disk $\sim 30\%$ and the thick disk (shown with upright triangles
in Figure~\ref{figs:surfb}) the remaining $\sim 10\%$.  The total
contribution of the (thick+thin) disk thus is of order $\sim 40\%$,
similar to that inferred from the photometric decomposition. This is
somewhat surprising, since the individual profiles deviate strongly
from the assumed $R^{1/4}$ and exponential laws. Indeed, neither
component can be adequately approximated by one such law; both show a
roughly exponential decay in the outer regions as well as a sharp
upturn near the center well described by a de Vaucouleurs' profile.

\begin{figure*}[t]
\epsscale{2.0}
\plottwo{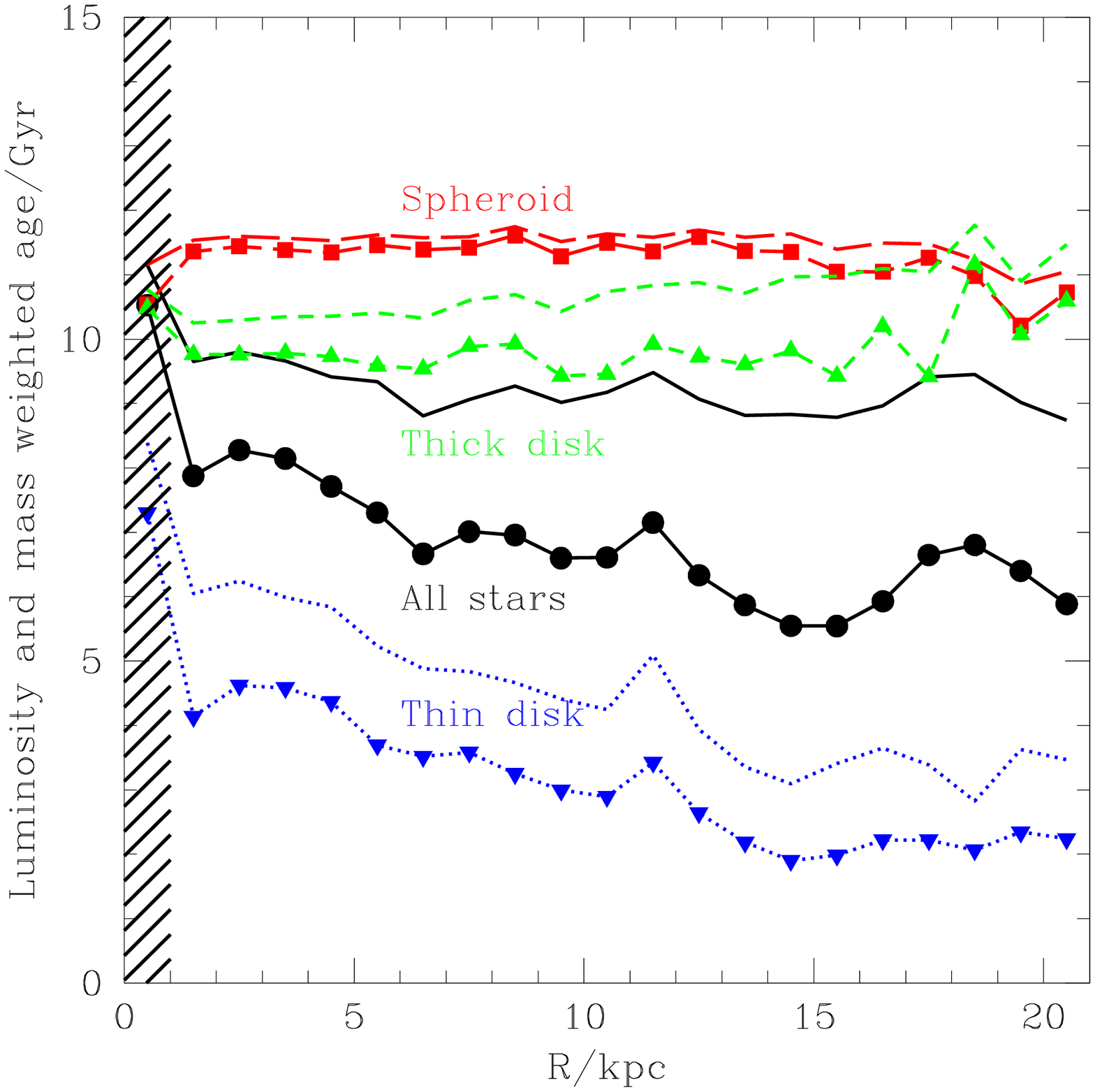}{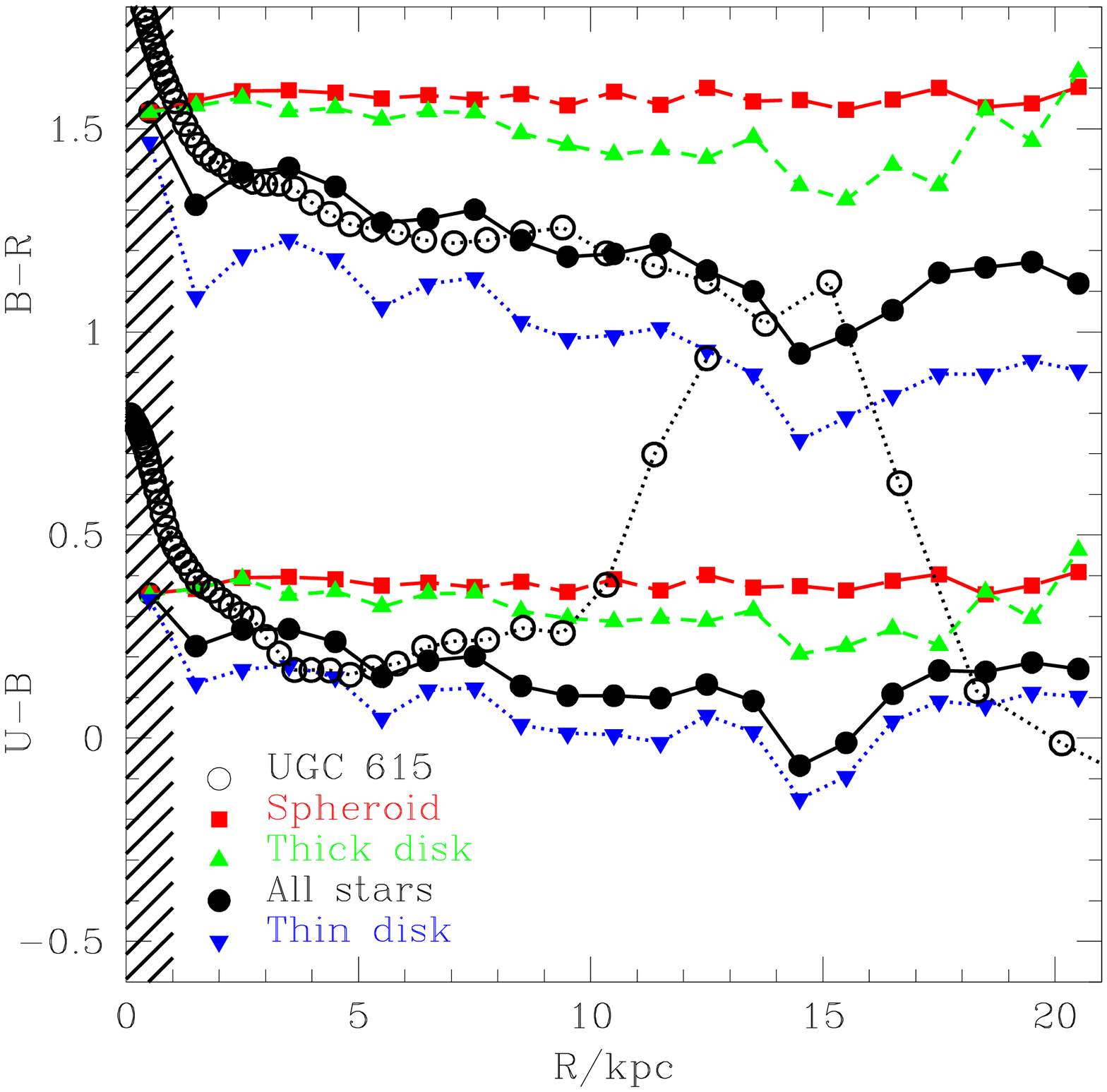}
\caption{(a) Left panel: Age profile of stars in the various components of the
simulated galaxy, seen face on. Curves connecting symbols correspond to
luminosity-weighted ages, those without symbols to mass-weighted
ages. Luminosities are computed assuming a Scalo IMF with upper and lower mass
cutoff of $100$ and $0.1 \, M_{\odot}$, respectively. (b) Right panel: $U-B$ and
$B-R$ color profiles of various components of the simulated galaxy seen face on
(filled symbols), and compared with the Sab galaxy UGC615 (open symbols).}
\label{figs:agecol}
\end{figure*}

The contribution of the thin disk to the total stellar mass of the
system is only $17\%$; its prominence in the surface brightness
profile is due to its younger age (and thus lower mass-to-light ratio)
compared with the spheroid. This is shown in Figure~\ref{figs:agecol};
the mean luminosity-(mass) weighted age of the thin disk is $3.6$
($5.3$) Gyr; a full $7.3$ ($6.0$) Gyr younger than the spheroid. For a
Scalo IMF (used throughout this analysis), this age difference results
in the $I$-band mass-to-light ratio of the thin disk being $\sim 2.2$
times brighter than the spheroid, accounting for the importance of the
disk in the photometric decomposition.  The thick disk is of
intermediate age between the thin disk and the spheroid; its
luminosity (mass)-weighted age is $9.8$ ($10.5$) Gyr.

Clearly, inferring the structural properties of the disk and
spheroidal components from photometric decomposition alone is not
straightforward. This may have important implications for studies that
attempt to assess the relative importance of cold,
centrifugally-supported disks and of hot, ``pressure''-supported
spheroids through photometric decomposition of large samples of
galaxies (Schechter \& Dressler 1987, Benson, Frenk \& Sharples,
2002). In particular, our study shows that it is critical to account
carefully for the age difference between components and its effect on
the mass-to-light ratios: assuming similar mass-to-light ratios for
the disk and spheroid would result in dramatic overestimation of the
dynamical importance of the disk component.

The results shown in Figure~\ref{figs:surfb} have repercussions as well for
studies that rely on photometric decomposition to gain insight into the process
of bulge formation. For example, the tight link between the spatial scalelengths
of the spheroidal and disk components derived photometrically might not
necessarily reflect a physical association between the two, as is usually
interpreted in models that view the formation of the bulge as a result of
secular evolution in the disk (MacArthur, Courteau, \& Holtzman 2002, Courteau,
de Jong, \& Broeils 1996).  For example, the half-mass radii of the thin disk
and spheroid are $\sim 5.0$ and $\sim 0.7$ kpc, respectively, which should be
contrasted with the $\sim 7.8$ and $\sim 1.1$ kpc that would be obtained from
the best $I$-band $R^{1/4}$+exponential fit assuming constant mass-to-light
ratios for each component.
 
Inferring the radial extent of the disk from photometric data alone thus leads
to a $\sim 50\%$ overestimate of the half-mass radius. This is an important
correction, since disk scalelengths estimated photometrically are usually used
to estimate the angular momentum of galaxy disks (see Paper I and references
therein).  It is important that angular momenta computed from numerical
simulations are estimated in the same way as observations in order to ensure
that any discrepancies reflect genuine differences in angular momentum content
and not systematic uncertainties in the estimation procedure.

With only one example, it would be premature to validate (or invalidate) the
conclusions of the studies mentioned above, but clearly caution is highly
recommended when using photometric data alone to assess dynamical theories for
the origin of the components of the galaxy. We intend to return to this question
when we have completed a more statistically significant sample of simulated
galaxies.

\epsscale{1.0}

\subsection{Dynamical Components and Stellar Populations} 
\label{ssec:stelpop}

The dynamical distinction at the basis of the decomposition procedure
described in \S~\ref{ssec:decomp} is closely linked to the time and
mode of assembly of each component. As shown in
Figure~\ref{figs:agecol}a, the spheroid is old and very well mixed
dynamically; its mass-weighted and luminosity-weighted ages are quite
uniform across the galaxy, with little sign of a radial gradient. The
thin disk, on the other hand, is fairly young and has a very well
defined radial gradient as a result of its inside-out assembly; the
luminosity (mass) weighted age drops from $\sim 5$ ($6$) Gyr at $R\sim
1$ kpc to just about $2$ ($3$) Gyr at $R=20$ kpc. The age difference
between components is reflected in their colors, as shown in
Figure~\ref{figs:agecol}b. The color profile of the simulated galaxy
as a whole (solid circles) shows a mild gradient, becoming bluer
farther from the center, in good agreement with that of the Sab galaxy
UGC615 (open circles). This trend is due almost entirely to the age
gradient in the thin disk component shown in panel (a) of
Figure~\ref{figs:agecol}; metallicity plays only a minor role in this
result.

Our results confirm two of the basic expectations of galaxy models based on
hierarchical accretion. (i) Spheroids are predominantly old and very well mixed
dynamically by merger events which weaken any preexisting age or metallicity
gradient in the progenitors. This is, of course, consistent with the Searle \&
Zinn (1978) interpretation of the lack of obvious gradients in the stellar halo
of the Milky Way as due primarily to its formation through merging of
protogalactic fragments. (ii) Centrifugally-supported disks, on the other hand,
are built from the inside out and are on average much younger than
spheroids. These results seem broadly consistent with observational data for
external galaxies and for our own Milky Way. We turn next to the detailed
dynamical origin of each component and to the subtle but important distinction
between the thin and the thick disk components.

\subsection {The Thin and Thick Disks}
\label{ssec:thnthck}

\begin{figure}[t]
\plotone{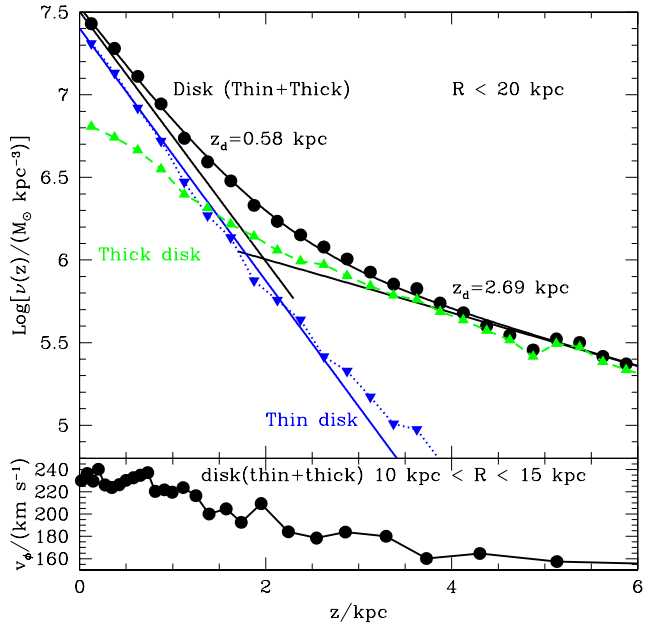}
\caption{Vertical density profile of disk (thin+thick) stars (solid
circles). A double exponential profile fits the profile very well,
with scalelengths of $575$ pc and $2.7$ kpc respectively. The
contribution of the dynamically-identified thin and thick disks, shown
by the lines connecting filled triangles are seen to agree fairly well
with those expected from the double exponential fit. The bottom inset
shows that the mean rotation speed of the disk decreases steadily with
increasing vertical distance from the plane of the disk. This shows
the gradual transition from the thin disk to the thick disk
component.}
\label{figs:nuzvrotz}
\end{figure}

We have identified a ``thick disk'' component with stars on
co-rotating (but rather eccentric) orbits which do not belong to
either the thin disk or the spheroid, according to the decomposition
procedure outlined in \S~\ref{ssec:decomp}. The combined thin+thick
disk component (hereafter ``the disk'', for short) resembles that of
the Milky Way in several respects. Consider, for example, the vertical
distribution of disk stars, shown in Figure~\ref{figs:nuzvrotz}. The
$z$-density profile deviates clearly from a simple exponential law,
but is well approximated by the combination of two exponentials, with
$575$ pc and $2.7$ kpc scaleheights respectively (see filled circles
and solid lines in Figure~\ref{figs:nuzvrotz}).

Interestingly, stars identified dynamically as belonging to the thin
disk follow closely an exponential profile as well, with scaleheight
$\sim 570$ pc.  The density profile of the thick disk is not exactly
exponential, but the median height of its stars is not very different
from that inferred from the double-exponential fit. The disk rotates
significantly slower at higher $z$, as shown in the bottom panel of
Figure~\ref{figs:nuzvrotz}. The mean rotation speed in the outskirts
of the disk ($10$ kpc < R < $15$ kpc, where the rotation curve is
approximately flat, see Paper I) drops from $240$ km/s on the plane of
the disk to $\sim 160$ km/s at $z \sim 3$ kpc.

Are the thick and thin disks really distinct components, or is the
double exponential profile shown in Figure~\ref{figs:nuzvrotz} just a
manifestation of a rotationally-supported structure intrinsically more
complex than a simple exponential? To be meaningful, a dynamical
decomposition between thick and thin disk components must also
identify other independent attributes, such as age or metallicity,
which neatly divide the stellar population into similar components.
In the case of the Milky Way, the metallicity and age distributions of
disk stars support the idea that the thin and thick disks are truly
distinct components of independent origin. Besides having higher
velocity dispersion, thick disk stars are typically older and more
metal poor than thin disk stars, and appear to divide ``naturally''
into two components in the age/metallicity plane (see, e.g., \S 10.4.3
of Binney \& Merrifield 1998).

\begin{figure}[t]
\plotone{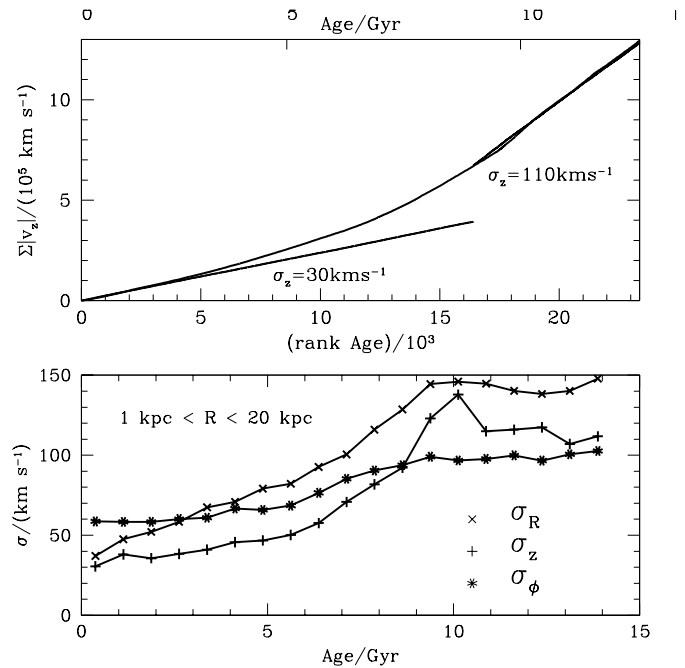}
\caption{The clear dynamical separation between thin and thick disks
is illustrated in the top panel by plotting the age rank of disk
particles versus the cumulative sum of the absolute value of the
vertical velocity component. Gaussian velocity distributions show as
straight lines with varying slope, depending on their velocity
dispersions. Two separate components, with $\sigma_z=30$ and $110$ km
s$^{-1}$, are clearly seen. The colder component is young (age $\lsim
8$ Gyr), whereas the hotter component is made up primarily of older
stars. The bottom panel shows that a similar conclusion is suggested
by the age-dependence of the velocity dispersion of disk stars.  }
\label{figs:sgagerank}
\end{figure}

The distinction between thin and thick disks is also often illustrated by
simultaneously ranking stars by their metal abundance and by their vertical
speed (see Freeman 1991). Since our treatment of metal enrichment is rather
rudimentary (\S~\ref{ssec:code}), we illustrate the same idea using age rather
than metallicity as one of the ranking variables. The break at $\sim 8$ Gyr seen
in the top panel of Figure~\ref{figs:sgagerank} is highly suggestive of the
presence of two components of distinct nature and origin: one young and cold,
well approximated by a velocity dispersion of $\sigma_z\approx 30$ km s$^{-1}$,
overlapping an older population of stars with a velocity dispersion roughly
three times as high. (A Gaussian distribution of velocities would show as a
straight line in this plot.)

This interpretation is confirmed by the velocity dispersion-age
relation for disk stars shown in the bottom panel of
Figure~\ref{figs:sgagerank}, where a sharp upturn in velocity
dispersion is noticeable for stars older than $\sim 8$ Gyr. These
results support the view that the thin and thick disks are truly
distinct components, and point to a major occurrence in the assembly
of the galaxy roughly $\sim 8$ Gyr ago: this timescale coincides with
the epoch of the last major accretion event.

\begin{figure}[t]
\plotone{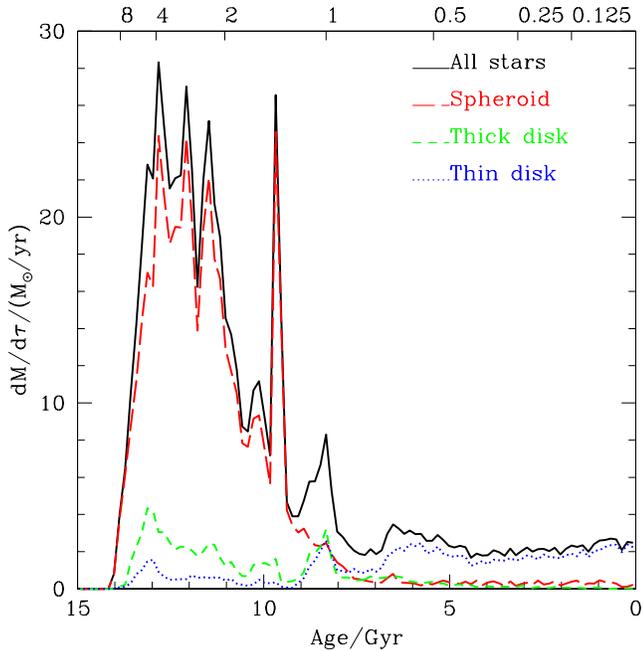}
\caption{Age distribution of stars in the galaxy (solid line), as well
as in its three dynamically-identified components: spheroid
(long-dashed lines), thick disk (short dashed line) and thin disk
(dotted line). Sharp peaks in the age distribution are typically
associated with starbursts triggered by merger and accretion
events. The last two starbursts are associated with a 1:2 merger at
$z=1.5$ and with the nearly simultaneous accretion of two satellites
at $z\sim 1$. Much of the accretion afterwards is in the form of
smooth settling of cooled gas, which is transformed into stars at the
roughly constant rate of $\sim 2\, M_{\odot}$/yr.}
\label{figs:agedist}
\end{figure}

Major accretion events usually trigger profuse star formation and thus
show clearly in the distribution of stellar ages shown in
Figure~\ref{figs:agedist}. For example, the sharp peak at age $\tau
\sim 9.5$ Gyr corresponds to a merger event which increases the
stellar mass of the galaxy by almost $50\%$. Similarly, the broad peak
at $\tau \sim 8$ Gyr is associated with the nearly simultaneous
accretion of two satellites; these and the associated starburst boost
the stellar mass of the galaxy by $\sim 20\%$. Subsequently, the main
galaxy is disturbed only by the accretion of two minor satellites (at
$z\sim 0.7$), each contributing less than $6\%$ of the current stellar
mass of the galaxy.  At lookback times less than $\sim 8$ Gyr most
stars form (and remain) in a thin disk as a result of the gradual
transformation into stars of cooled gas accreted smoothly into the
main body of the galaxy.

Clearly, the sharp decline in merger activity which occurred $\sim 8$ Gyr ago is
the watershed event responsible for initiating the formation of the thin disk
and for establishing the neat separation between the thin and thick disk
components. The bulk of thick disk stars is therefore older than $8$ Gyr and the
majority of thin disk stars form after that epoch. We explore next the origin of
the thick disk in the simulated galaxy and the applicability of these findings
to the formation of the Milky Way's thick disk.

\subsection{The Origin of the Thick Disk}
\label{ssec:thkdsk}

As discussed in \S~\ref{sec:intro}, dynamical data in the solar neighborhood are
usually interpreted within a scenario where the thick disk results from the
tidal heating of an early thin stellar disk. Does the thick disk in the
simulated galaxy form in this way? We can test this proposal directly by tracing
the stars assigned to the thick disk at $z=0$ and by inspecting their
configuration at the time of the last major merger.

The result of this exercise is summarized in the bottom panel of
Figure~\ref{figs:finsitu}. This figure shows, as a function of age, the mass
fraction of each component which formed within the main progenitor of the
simulated galaxy.  The evolution of this ``in situ''-formed mass fraction is
similar for the thin and thick disks, and indicate that essentially all disk
stars younger than $\sim 8$ Gyr were formed within the main galaxy, a result
consistent with the lack of major accretion events since $z\sim 1$.

On the other hand, the majority of old disk stars were formed in systems other
than the main progenitor; only about $10\%$ of disk stars older than $\sim 10$
Gyr were formed in the main progenitor of the galaxy. {\it This implies that the
bulk of old disk stars are actually tidal debris from disrupted
satellites}. Since the thick disk is quite old (only $\sim 12\%$ of its stars
are younger than $8$ Gyr, see the top panel of Figure~\ref{figs:finsitu}), this
rules out the hypothesis that the thick disk is the remnant of a stirred
pre-existing thin disk and points to tidal disruption as the main process of
assembly of the thick disk.  About $\sim 70\%$ of thick disk stars older than
$8$ Gyr are debris from tidally disrupted satellites; in total, more than $60\%$
of thick disk stars of all ages were brought in by satellites, the rest were
mostly formed during the accretion events themselves.

Interestingly, although the bulk of the thick disk consists of debris from
tidally disrupted satellites, the fraction of spheroid stars identified as tidal
debris is only $\sim 50\%$ of the total. The other half of the stars in the
spheroid formed in a disk-like component in the main galaxy but were dispersed
into the spheroid as a result of a major ($1$:$2$) merger at $z\approx 1.5$
($\sim 9.5$ Gyr ago). Thus, our results suggest that, contrary to what might be
naively expected, it would be easier to find remnants of tidal stripping events
in the old disk component rather than in the spheroid.

Considering the thick disk as the accumulated tidal debris of
disrupted satellites rather than as an early thin disk thickened by an
accretion event helps to resolve the angular momentum puzzle alluded
to in \S~\ref{sec:intro}. {\it The high angular momentum content of
the thick disk does not reflect the presence of a large disk at high
redshift but rather the orbital angular momentum of disrupted
satellites, some of which were aggregated into the main galaxy
relatively recently.} In our simulation, for example, half of the
stars in the thick disk are older than $\sim 10$ Gyr, but were fully
accreted into the main progenitor only $\sim 6$ Gyr ago. We explore
below how this result also affects the interpretation of very old
stars on circular orbits in the outskirts of the disk galaxies.

\subsection{The Old Thin Disk}
\label{ssec:oldthndsk}

\begin{figure}[t]
\plotone{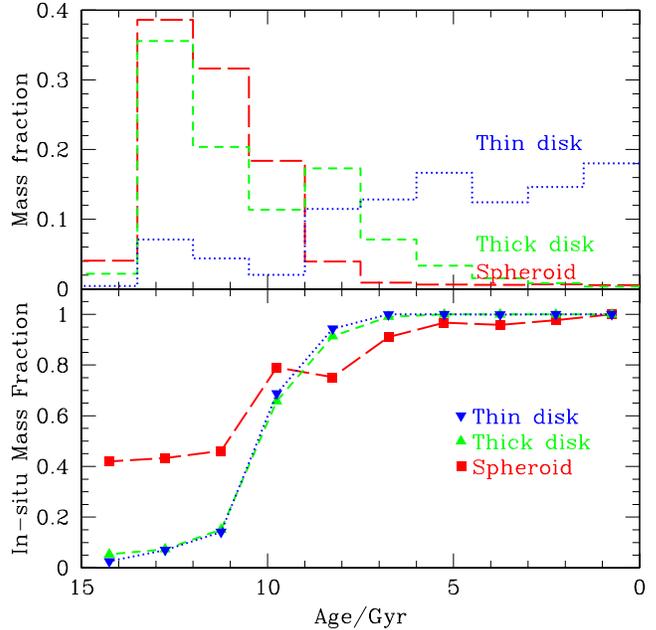}
\caption{Top panel: Stellar age distribution (as in
Figure~\ref{figs:agedist}), but normalized to the total mass of each
individual component.  The spheroid and the thick disk are quite old;
the thin disk is relatively young. However, $\sim 15\%$ of the thin
disk stars are older than $\sim 10$ Gyr, and thus {\it predate} the
last few major mergers. Bottom panel: Fraction of stars in each of the
bins of the top panel formed within the main galaxy (``in situ''), as
a function of time. Note that most of the old ($\tau \gsim 10$ Gyr)
disk (thick or thin) was {\it not} formed in situ, and thus is made up
of the debris of tidally dispersed satellites.}
\label{figs:finsitu}
\end{figure}

As discussed in \S~\ref{sec:intro}, the existence of disk stars which predate
the epoch of the last major merger but which are nevertheless found today on
nearly circular orbits is puzzling in hierarchical assembly models of disk
formation. Debris from tidally disrupted satellites offers again a plausible
explanation for the presence of such a component. The simulated galaxy has a
well defined ``old thin disk'' population; indeed, Figure~\ref{figs:finsitu}
shows that $\sim 15\%$ of the thin disk stars are older than $10$ Gyr, and
therefore formed long before the time of the last major accretion event. As
shown in the bottom panel of the same Figure, essentially all of these stars
were formed in satellites and brought into the main galaxy much later; the last
of these ``old thin disk'' stars was only incorporated into the main galaxy
roughly $6$ Gyr ago.

\begin{figure*}[t]
\centerline{\epsscale{1.65}}
\plotone{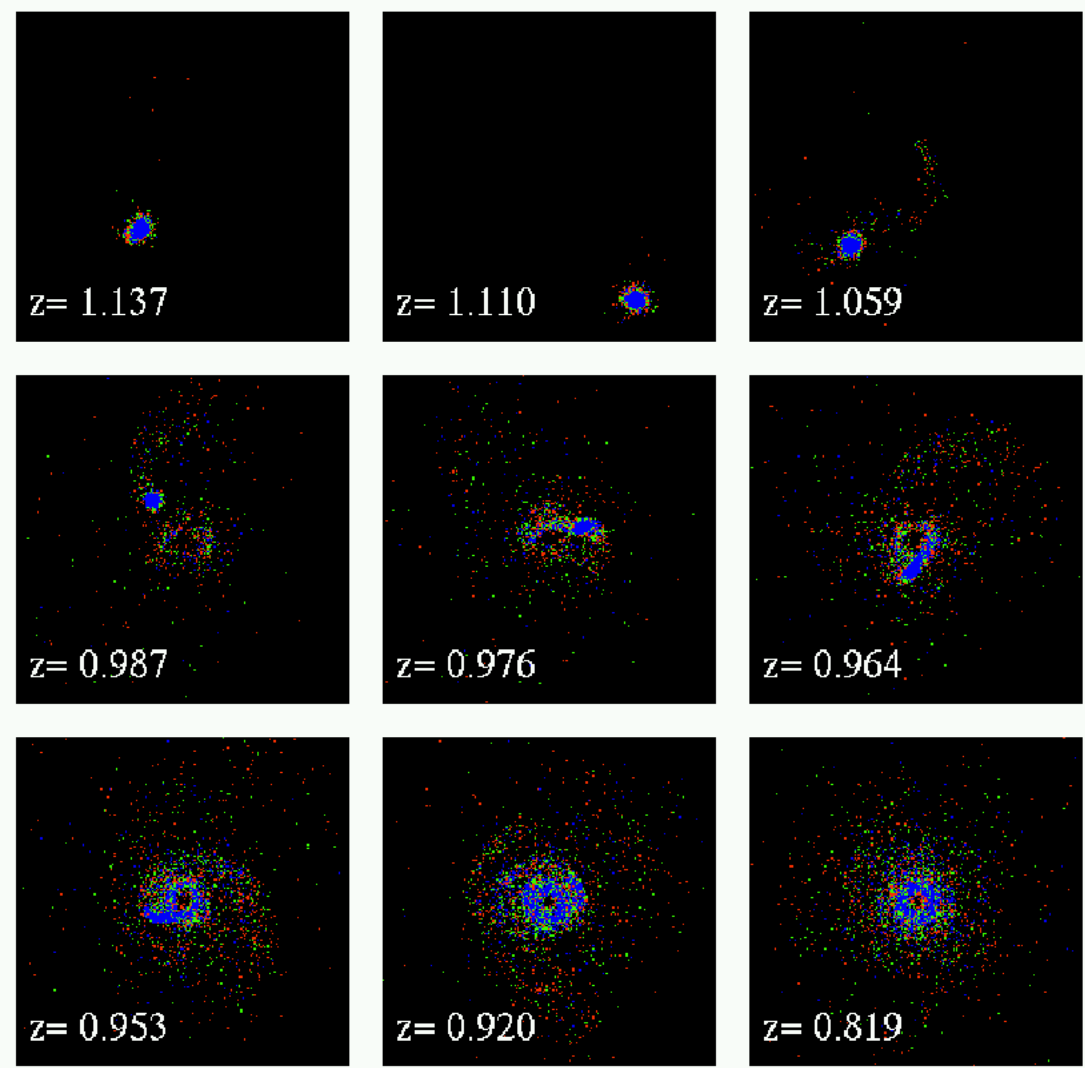}
\caption{Several stages in the disruption of satellite 1 (see
Table~\ref{tab:satprop}). The main galaxy (not shown) is at the center
of each panel. This satellite contributes a fair fraction ($\sim
77\%$) of its stars to the disk, and is seen projected face-on in this
figure and edge-on in Figure~\ref{figs:sat1y}. Particles are colored
according to their orbital circularity at $z=0$. The core of the
satellite (shown in blue) ends up contributing mainly to the thin
disk; the outskirts contribute to the thick disk (green) and to the
spheroid (red). The orbit of the satellite is brought into the disk
and circularized by dynamical friction until its final
disruption. Note that most stars in the satellites are dispersed into
a torus-like structure of radius given by the distance from the center
at which final disruption takes place.}
\label{figs:sat1z}
\end{figure*}

Figure~\ref{figs:finsitu} thus demonstrates that tidal debris from satellites
can contribute not only to the spheroid (as is usually assumed) but also to the
thin and thick disk components. A satellite's contribution to either component
(or to the spheroid) depends mainly on its orbit and on the degree to which
dynamical friction circularizes the orbit before disruption. In order to
contribute significantly to the disk, a satellite must have an orbital plane
roughly coincident with that of the disk and be dense enough to survive
disruption until its orbit has been circularized within the disk.

This intuitive expectation is confirmed by the data presented in
Table~\ref{tab:satprop}, where we list the structural properties and
orbital parameters of four different satellites accreted after $z\sim
1$. This Table lists the satellite's orbital circularity, as well as
the fraction of stars that end up, at $z=0$, in the spheroid, thin
disk, and thick disk components, respectively. Since the orbital
properties change significantly with time, we have chosen to measure
them at $z_{\rm acc}$, the time of their last apocentric passage
before significant disruption takes place. 


As expected, satellites on orbits of high circularity are seen to
contribute principally to the disk component rather than to the
spheroid, and viceversa. Closer inspection shows that it is the
satellite's core that contributes primarily to the thin disk whilst
stars in its outskirts end up primarily in the thick disk or spheroid
(see Figure~\ref{figs:sat1z}). This is fully consistent with the
notion that dynamical friction circularizes the orbit of the satellite
prior to disruption: stars at the core of the satellite are more
resilient to disruption and therefore end up on more circular orbits
than those stripped earlier during the satellite decay process.

\begin{figure*}[t]
\centerline{\epsscale{1.65}}
\plotone{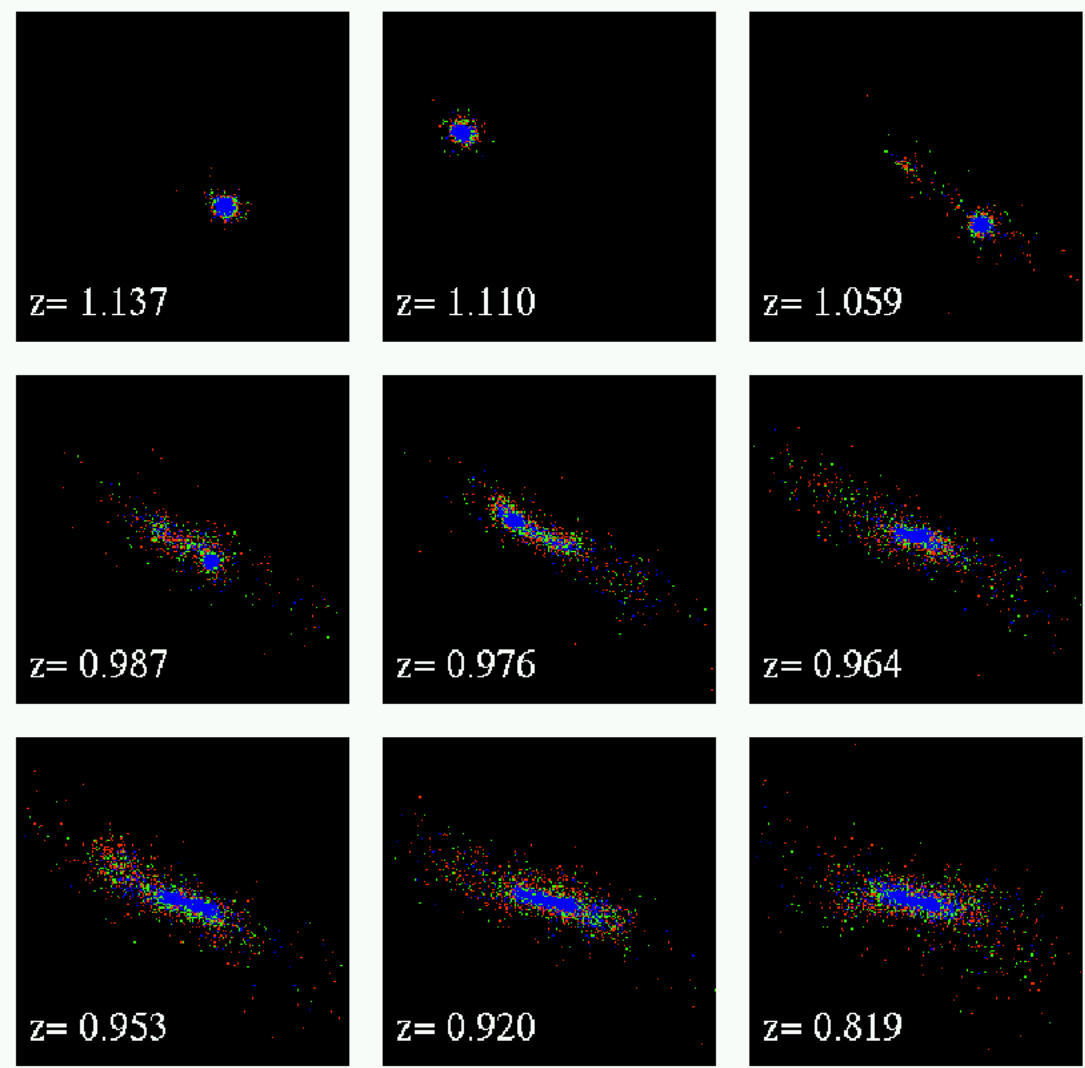}
\caption{As in Figure~\ref{figs:sat1z}, but seen edge-on.}
\label{figs:sat1y}
\end{figure*}
\epsscale{1.0}

The discussion above implies that, although the disk component is
formed overall from the inside out (see, e.g.,
Figure~\ref{figs:agecol}), a small fraction of old stars on circular
orbits is expected at essentially all radii as a result of satellite
accretion. Is this a result of general applicability or just a fluke
in this simulation? The answer depends on the incidence of capture of
satellites on orbital planes roughly coincident with that of the disk
of the galaxy.

With the obvious caveat that it is perilous to generalize from a
single example, there is indication that the accretion of such
satellites might not be uncommon in hierarchical scenarios. As
Table~\ref{tab:satprop} indicates, all $4$ satellites accreted by the
galaxy since $z\sim 1$ co-rotate with the disk and, after disruption,
contribute roughly $\sim 40\%$ of their stars to the disk. This
result, as well as the relatively high frequency of satellites with
orbital planes coincident with the disk, are apparently related to the
fact that much of the late accreting material (which effectively forms
the disk) flows in through a filament.

This filament inevitably contains previously collapsed, condensed
subsystems; satellites whose orbits share the properties of the
material responsible for the formation of the disk. If dense enough,
these satellites survive disruption until after their orbits are
circularized and contribute a substantial fraction of their stars to
the disk. We expect this to be a common feature of the hierarchical
formation process of a disk galaxy, implying that the majority of
galactic disks ought to have a significant population of old stars on
orbits similar to those of young disk stars but with distinct
properties such as age and metallicity. We intend to explore fully the
consequences of this idea in future papers of this series. We discuss
briefly below the implications of these findings for the Milky Way
galaxy.

\begin{figure*}[t]
\epsscale{2.2}
\plottwo{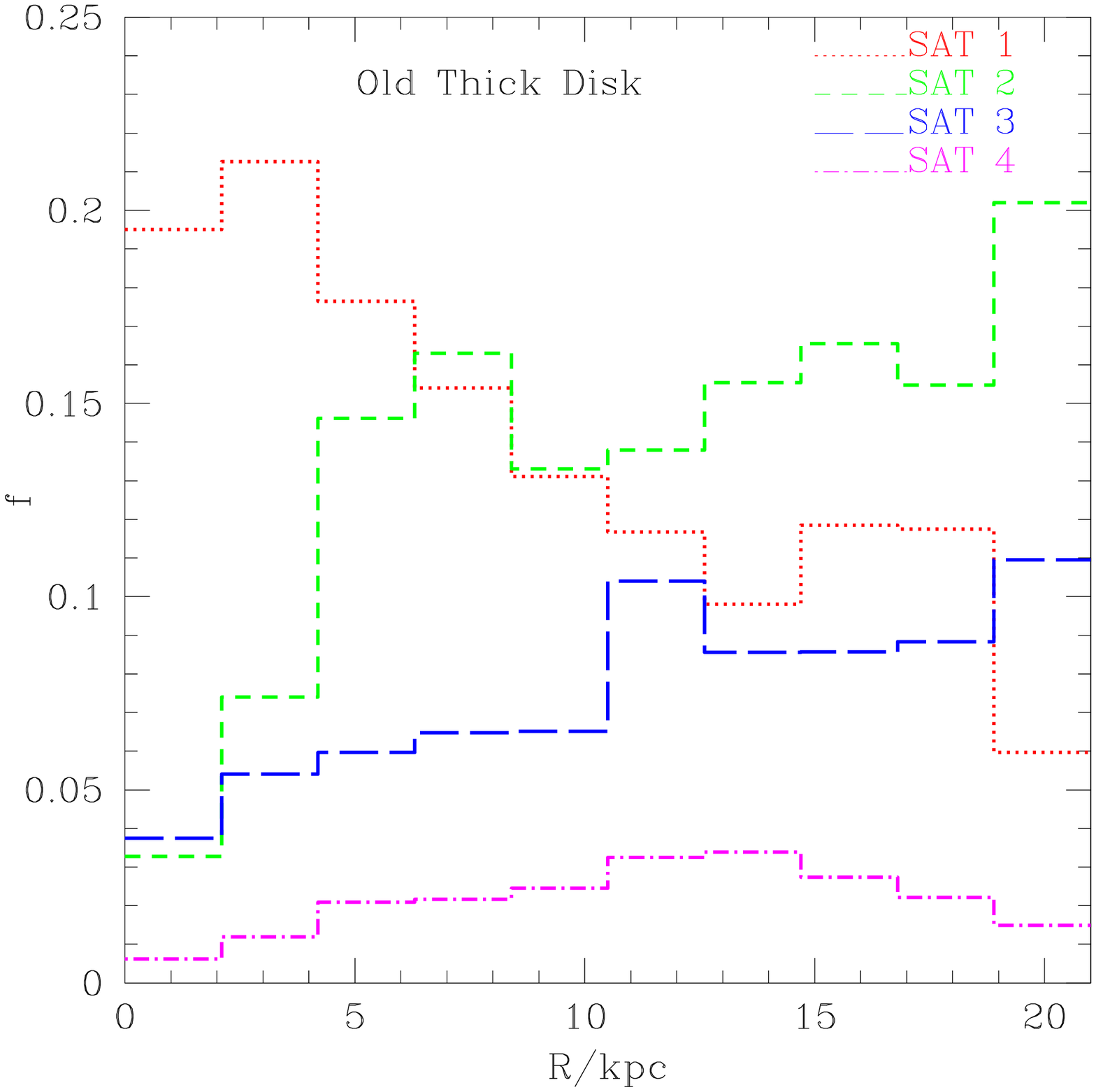}{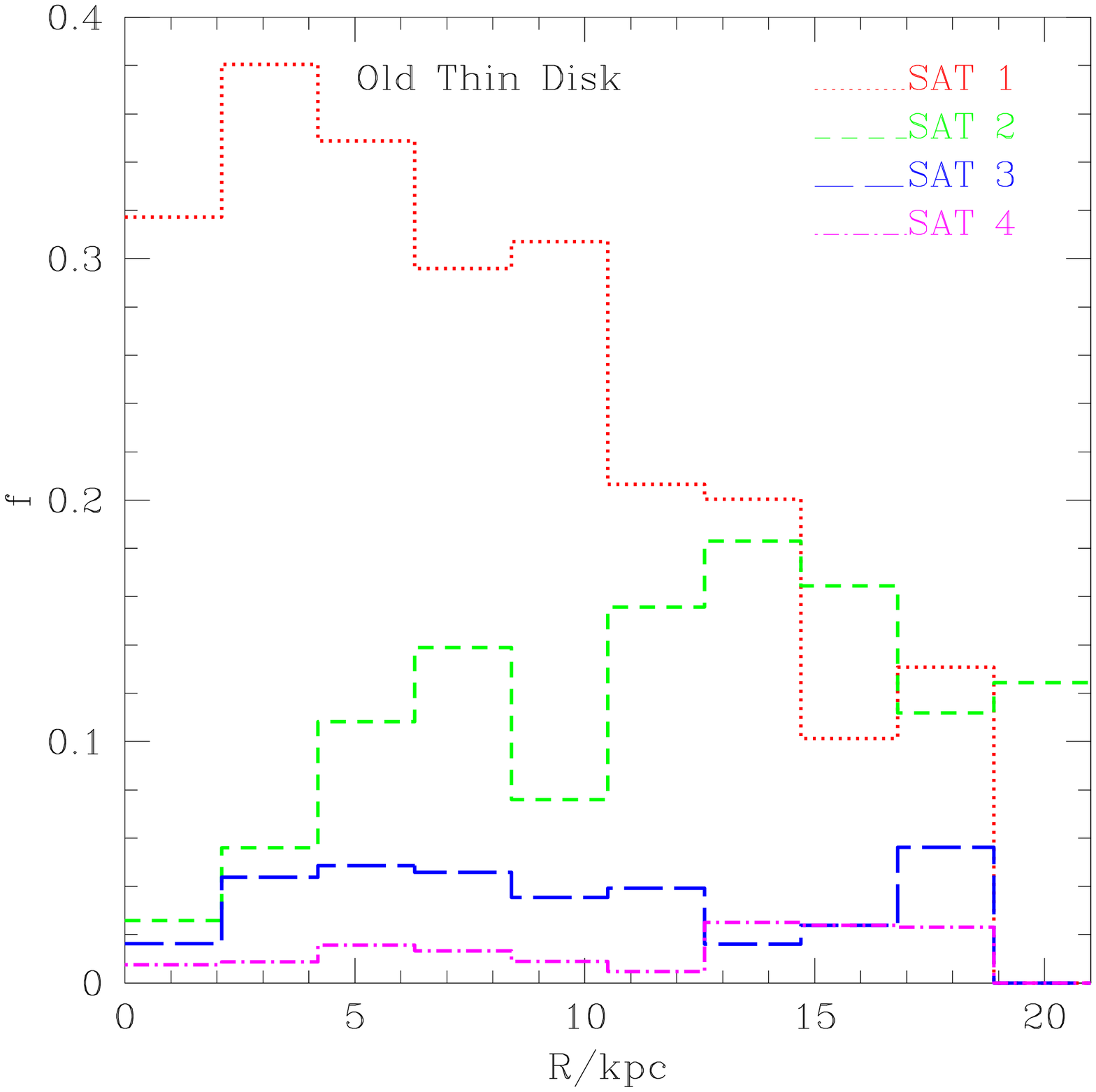}
\caption{ (a) Left panel: The fraction of stars contributed by
satellites 1-4 (see Table~\ref{tab:satprop}) to the old ($\tau>10$
Gyr) thick disk component, as a function of radius. (b) Right panel:
same as left panel, but for the old thin disk component.
\label{figs:satprof}}
\end{figure*}
\epsscale{1.0}

\subsection{Implications for the Milky Way}
\label{ssec:mwimp}

The discussion above illustrates the important clues to the accretion history of
the galaxy retained in the detailed dynamics and structure of the oldest stellar
populations. In particular, it shows conclusively that relicts of past accretion
and disruption events might be found today not only in the stellar halo of the
Galaxy, as is commonly assumed (Helmi et al. 1999, Majewski et al 2000), but
also in the cold, centrifugally supported disk component. As mentioned above,
tidal debris is proportionally more abundant in the old disk component of the
simulated galaxy than in the old spheroid; $\sim 85\%$ of disk stars older than
$10$ Gyr are the relicts of disrupted satellites, well in excess of the $\sim
55\%$ of the old stellar spheroid contributed by satellite debris. This suggests
that it would be profitable to search for signatures of accretion events in the
disk as well as in the halo components; indeed, such relicts may have already
been detected in the unexpected dynamics of stars above and below the plane of
the Galaxy reported by Gilmore et al (2002).

Could the ``old disk'' in the Milky Way be made up of the remains of disrupted
satellites? It is important to verify that this proposal is consistent with the
observation that the radial distribution of old disk stars is not very different
from that of the younger thin disk component. This is especially true because
stars from a single disrupted satellite are not uniformly mixed radially but
rather form a torus-like structure whose radius is determined by the location at
which final disruption occurs. This is clearly shown in Figure~\ref{figs:sat1z},
where the stars of satellite $1$ (see Table~\ref{tab:satprop}) are shown at
different times during its disruption. Once the disruption is complete, the
satellite stars are distributed along a thin ``ring'' of stars at a radius $\sim
1.7$ kpc from the center.

On the other hand, several satellites contribute to the old disk
component, and each contributes in varying amounts at different
radii. This is shown in the left panel of Figure~\ref{figs:satprof};
the debris from satellite $1$ dominates the inner regions ($R\lsim 5$
kpc) of the old thick disk, whereas debris from satellite $2$ is the
principal contributor to the outer regions. Satellites $3$ and $4$
contribute fewer stars at all radii because their orbits are far less
circular (see Table~\ref{tab:satprop}). 

Fewer satellites are likely to contribute to the thin disk than to the thick
disk. This implies that, at given radius, the old thin disk might be dominated
by the contribution of a single satellite. This is indeed the case in the
simulated galaxy at $R\lsim 10$ kpc, as shown in
Figure~\ref{figs:satprof}b. Systematic variations in the properties of old disk
stars as a function of their rotational support are thus expected and should
contain intriguing clues to the early assembly process of the disk. This subtle
modulation might be observable in our own Milky Way, once surveys such as those
planned for the space-based GAIA mission (Perryman et al 2001) and the ground-based RAVE
project (Steinmetz 2002)---where full dynamical information will be available
for large samples of stars well beyond the solar neighborhood---are completed.

\begin{figure}[t]
\plotone{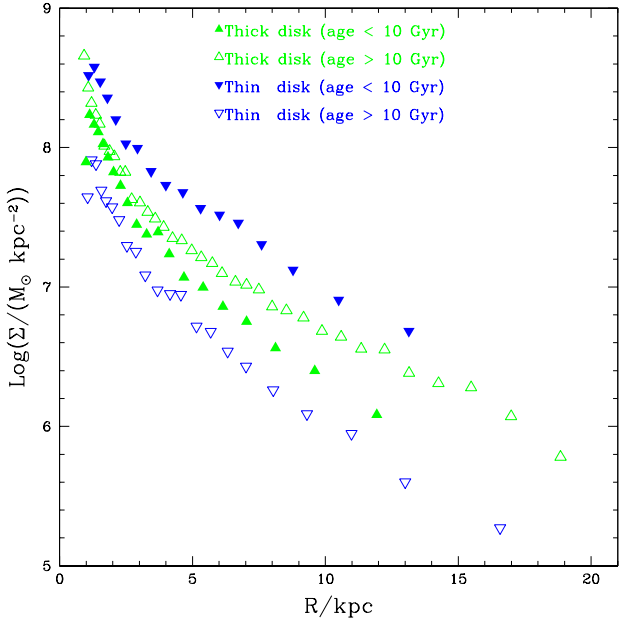}
\caption{Stellar surface mass density profile of the disk, split into
four components; thick/thin and young/old, where we term as ``old''
stars of ages exceeding $9$ Gyr.  All four components have similar
half-mass radii, and thus there is not a very substantial difference
in the radial distribution of disk stars, whether segregated by age or
by orbital circularity.}
\label{figs:binsurf_dsk}
\end{figure}

Overall, the radial overlap of debris from several disruption events results in
a radial distribution of old disk particles that differs little from that of the
young disk component, in agreement with observation. This is illustrated in
Figure~\ref{figs:binsurf_dsk}, where we show the surface density profile of the
``young'' and ``old'' disk components. There are no major differences in the
radial distribution of the old and young disks; the half-mass radius of the old
thin (thick) disk is $3.9$ ($5.6$) kpc, comparable to that of the young thin
(thick) disk, $5.2$ ($3.9$) kpc, despite the large difference in the mean age of
each component. {\it The late accretion of satellites with orbits coincident
with the disk thus explains the puzzle alluded to in \S~\ref{sec:intro}
regarding the radial extent (and angular momentum content) of the older, thicker
disk component of the Galaxy.}

If much of the old thick (and thin) disk is indeed debris
from tidally disrupted satellites, their metallicities must reflect
the metallicity of the satellites themselves. Since metallicity and
luminosity are strongly correlated in dwarf systems, then one may use
the metallicity of the old disk component to infer the luminosity of
the disrupted satellite. For example, if the metallicity of the thick
disk is taken to be [Fe/H]$\approx -0.6$, one would infer that the
luminosity of its progenitor satellite had a luminosity comparable to
that of the LMC, $M_V\sim -18.5$ (see Figure 8 of van den Bergh
1999).

Another important implication of this idea is that the {\it same} satellite can
contribute both to the old thin and thick disks. This implies that these two
components should exhibit a clear overlap in properties, reflecting their common
origin in a single system. This is reminiscent of the results of Wyse \& Gilmore
(1995), who argue that the dynamical and metallicity distributions of disk stars
can only be explained if stars with abundances [Fe/H] $<-0.4$ contribute in
substantial amounts to both the thick and thin disk components.


We end this discussion with the obvious caveat that, although we find
the formation mechanism of the old disk proposed here attractive and
applicable to the Milky Way, the simulated galaxy is {\it not}
intended to be a model of the Galaxy. Further simulation work is
required to ascertain the general validity of the results reported
here and their applicability to the Milky Way and other spirals.

\section{Summary}
\label{sec:conc}

We present a detailed analysis of the dynamical components of a galaxy
simulated in the ``concordance'' $\Lambda$CDM cosmogony. The galaxy
forms in a dark matter halo chosen so that mergers and accretion
events are unimportant dynamically after $z\sim 1$. Star formation and
feedback parameters are such that the star formation history of the
galaxy is largely driven by the rate at which gas cools and collapses
within dark halos: this is a conservative choice where feedback
effects are relatively unimportant. The shortcomings of this
assumption concerning the global photometric and structural properties
of the simulated galaxy are discussed in Paper I. We focus here on the
multi-component nature of the stellar disk and on their dynamical
origin.  Our main results can be summarized as follows.

\begin{itemize}

\item{At $z=0$, two separate stellar components are easily
distinguishable solely on the basis of the orbital parameters of stars
in the galaxy: a slowly rotating, centrally concentrated spheroid and
an extended disk-like component largely supported by rotation. }

\item{These components are also recognized in the surface
brightness profile of the galaxy, which can be very well approximated by the
superposition of an $R^{1/4}$ spheroid and an exponential disk, in agreement
with observations of early-type spirals.}

\item{ Neither does the dynamically identified spheroid follow closely
de Vaucouleurs' law nor is the disk purely exponential, a result which
calls for caution in the interpretation of estimates of the dynamical
importance of the disk and spheroid in traditional photometric
decomposition techniques.}

\item{The spheroid is old, and has essentially no stars younger than
the last major accretion episode; $\sim 8$ Gyr ago for the system we
consider here. The majority of thin disk stars, on the other hand,
form after the merging activity is over and have a mean age of $\sim 5$
Gyr.}

\item{The disk may be further decomposed into two well defined
subcomponents: a thin, dynamically cold disk of stars on nearly circular orbits
and a thicker disk with orbital parameters transitional between the thin disk
and the spheroid.  Supporting evidence for the true presence of these two
distinct components is found, as in the Milky Way, in the double-exponential
vertical structure of the disk and in abrupt changes in the vertical velocity
distribution as a function of age.}

\item{The bulk ($\sim 60\%$) of the thick disk consists of the tidal
debris of satellites whose orbital plane was coincident with the disk
and whose orbits were circularized by dynamical friction prior
to full disruption. This trend becomes more pronounced with age; $\sim
90\%$ of stars older than $10$ Gyr and presently in the thick disk
component were brought into the disk by satellites.}

\item{ A significant fraction ($\sim 15\%$) of thin-disk stars are old
enough ($\gsim \, 10$ Gyr) to predate the last major accretion
event. The bulk of this unexpected population of old stars on nearly
circular orbits share a common origin with the old thick disk: they
are the remains of the cores of disrupted satellites. Interestingly,
only one in two stars belonging to the old spheroid are tidal debris;
the rest may be traced to a major merger event that disperses the
luminous progenitor at $z\sim 1.5$ and seeds the formation of the
spheroid.}

\end{itemize}


Our results offer clues to understanding a number of observational
trends that challenge the standard hierarchical disk assembly
process. In particular, the presence of an old disk component made up
primarily of tidal debris from disrupted satellites helps to explain:
(i) the presence of a significant number of old stars on circular
orbits in the outskirts of galaxies like the Milky Way, and (ii) why
the specific angular momentum and radial extent of the thick and thin
disk components are comparable in spite of the significantly different
ages of their individual stars. It also offers a ``natural''
explanation for the clear dynamical and evolutionary distinction
between the thin and thick disk components: the thick disk is mostly
tidal debris from disrupted satellites whereas the young thin disk
consists mostly of stars formed ``in situ'' after the merging activity
abates.

These findings highlight the role of satellite accretion events in
shaping the disk---as well as the spheroidal---components of a galaxy
and reveal some of the clues to the assembly process of the galaxy
preserved within the detailed dynamics of old stellar populations. We
emphasize that these conclusions are based on the detailed analysis of
a single simulation and that further work is needed to put these
results on a firmer basis as well as to establish their true
applicability to the origin of the dynamical components of the Milky
Way. We plan to explore these issues in detail in future papers of this
series.

\acknowledgments

This work has been supported by grants from the U.S. National
Aeronautics and Space Administration, the Natural Sciences and
Engineering Research Council of Canada, and Fundaci\'on Antorchas from
Argentina.

\clearpage

\begin{deluxetable}{lcccccccc}
\tablecaption{Properties of the stellar components of the galaxy\label{tab:gxprop}}
\tablehead{
\colhead{Label} &
\colhead{$N_{\rm part}$} & 
\colhead{$M_{\rm }$} & 
\colhead{$L_{\rm I}$} &
\colhead{$\Upsilon_I$} &
\colhead{$L_{\rm I}^{\rm fit}$} &
\colhead{$R_{\rm hm}$} & 
\colhead{$R_{\rm hl}^{\rm fit}$} &
\colhead{$\bar{\epsilon}_J$}  
\\ 
\colhead{} &
\colhead{} &
\colhead{[$10^{10}\, M_{\odot}$]} & 
\colhead{[$10^{10}\, L_{\odot}$]} & 
\colhead{[$M_{\odot}/L_{\odot}$]} & 
\colhead{[$10^{10}\, L_{\odot}$]} & 
\colhead{[kpc]} &
\colhead{[kpc]} &
\colhead{} 
\\ 
}
\startdata
spheroid   & $79423$ & $7.23$ & $2.63$ & $2.74$ & $2.33$ & $0.67$ & $1.12$ & $0.12$ \\
disk       & $27797$ & $3.01$ & $1.93$ & $1.56$ & $1.99$ & $4.99$ & $7.79$ & $0.80$ \\
thin disk  & $16406$ & $1.79$ & $1.43$ & $1.25$ & $ -  $ & $5.05$ & $ -  $ & $0.92$ \\
thick disk & $11391$ & $1.22$ & $0.50$ & $2.44$ & $ -  $ & $4.89$ & $ -  $ & $0.64$ \\
\enddata
\tablecomments{Summary of the properties of the dynamically-identified
components of the galaxy. All quantities are computed within $r_{\rm
lum}=21$ kpc. We note that not all star particles have the same
mass. Quantities labelled with a ``fit'' superscript are computed from
the photometric $R^{1/4}$+exponential best fit described in
\S~\ref{ssec:photdyn}. $R_{\rm hm}$ ($R_{\rm hl}$) denotes the
half-mass (light) radius of each component; $\bar{\epsilon}_J$ their
average circularity.}
\end{deluxetable}

\begin{deluxetable}{crcrcrcr}
\tablecaption{Properties of satellites accreted after $z\sim 1$.\label{tab:satprop}}
\tablehead{
\colhead{Label} &
\colhead{$M_{\rm stars}$} & 
\colhead{$R_{\rm hm}$} & 
\colhead{$z_{\rm acc}$} & 
\colhead{$\epsilon_J$} &
\colhead{$f_{\rm thn}$} & 
\colhead{$f_{\rm thk}$} &
\colhead{$f_{\rm spher}$} 
\\ 
\colhead{} &
\colhead{[$10^{10}\, M_{\odot}$]} & 
\colhead{[kpc]} &
\colhead{} &
\colhead{} &
\colhead{} &
\colhead{} &
\colhead{} 
\\ 
}
\startdata
$1$ & $0.42$ & $0.54$ & $1.03$ & $0.87$ & $0.33$ & $0.44$ & $0.23$ \\
$2$ & $0.27$ & $0.61$ & $0.76$ & $0.48$ & $0.11$ & $0.44$ & $0.45$ \\
$3$ & $0.57$ & $0.71$ & $0.73$ & $0.00$ & $0.01$ & $0.05$ & $0.94$ \\
$4$ & $0.10$ & $0.51$ & $1.06$ & $0.30$ & $0.02$ & $0.16$ & $0.82$ \\
\enddata
\tablecomments{Properties of four satellites accreted by the main
galaxy after $z\sim 1$. $R_{\rm hm}$ is the stellar half-mass radius,
and $\epsilon_J$ the orbital circularity measured at $z_{\rm acc}$,
the redshift of the last apocentric passage of the satellite before
significant disruption takes place. The parameters $f_{\rm thn}$,
$f_{\rm thk}$, and $f_{\rm spher}$ denote the fraction of stars that
end up, at $z=0$ in the thin disk, thick disk, or spheroidal component
of the galaxy, respectively. Note the good correlation between the
orbital circularity and the fraction of stars that contribute to the
disk component.}
\end{deluxetable}


\end{document}